\documentclass[
    aps,
    prd,
    reprint,
    superscriptaddress,
    amsmath,
    amssymb
]{revtex4-2}

\usepackage{graphicx}
\usepackage{bm}
\usepackage{makecell} 
\usepackage{hyperref}

\begin{document}

\author{D.\,D.\,Dzhappuev}
\author{I.\,M.\,Dzaparova}
\affiliation{Institute for Nuclear
Research of the Russian Academy of Sciences, 60th October Anniversary Prospect 7a, Moscow 117312, Russia}
\author{T.\,A.\,Dzhatdoev}
\affiliation{Institute for Nuclear
Research of the Russian Academy of Sciences, 60th October Anniversary Prospect 7a, Moscow 117312, Russia}
\affiliation{Lomonosov Moscow State University, 1-2 Leninskie Gory, Moscow 119991, Russia}
\author{E.\,A.\,Gorbacheva}
\author{I.\,S.\,Karpikov}
\author{M.\,M.\,Khadzhiev}
\author{A.\,U.\,Kudzhaev}
\author{A.\,N.\,Kurenya}
\author{A.\,S.\,Lidvansky}
\author{O.\,I.\,Mikhailova}
\affiliation{Institute for Nuclear
Research of the Russian Academy of Sciences, 60th October Anniversary Prospect 7a, Moscow 117312, Russia}
\author{V.\,B.\,Petkov}
\affiliation{Institute for Nuclear
Research of the Russian Academy of Sciences, 60th October Anniversary Prospect 7a, Moscow 117312, Russia}
\author{N.\,A.\,Pozdnukhov}
\email[Corresponding author; email: ]{nikita.pozdnukhov1@gmail.com}
\affiliation{Institute for Nuclear
Research of the Russian Academy of Sciences, 60th October Anniversary Prospect 7a, Moscow 117312, Russia}
\author{V.\,S.\,Romanenko}
\author{G.\,I.\,Rubtsov}
\affiliation{Institute for Nuclear
Research of the Russian Academy of Sciences, 60th October Anniversary Prospect 7a, Moscow 117312, Russia}
\author{S.\,V.\,Troitsky}
\affiliation{Institute for Nuclear
Research of the Russian Academy of Sciences, 60th October Anniversary Prospect 7a, Moscow 117312, Russia}
\affiliation{Lomonosov Moscow State University, 1-2 Leninskie Gory, Moscow 119991, Russia}
\author{I.\,B.\,Unatlokov}
\affiliation{Institute for Nuclear
Research of the Russian Academy of Sciences, 60th October Anniversary Prospect 7a, Moscow 117312, Russia}
\author{N.\,A.\,Vasiliev}
\affiliation{Institute for Nuclear
Research of the Russian Academy of Sciences, 60th October Anniversary Prospect 7a, Moscow 117312, Russia}
\affiliation{Lomonosov Moscow State University, 1-2 Leninskie Gory, Moscow 119991, Russia}
\author{A.\,F.\,Yanin}
\author{K.\,V.\,Zhuravleva}
\affiliation{Institute for Nuclear
Research of the Russian Academy of Sciences, 60th October Anniversary Prospect 7a, Moscow 117312, Russia}
\collaboration{Carpet--3 Group}
\noaffiliation

\title{Isotropic diffuse gamma-ray flux limits from Carpet-3 experiment data using neural networks with domain adaptation}

\begin{abstract}
    Diffuse gamma rays above $100$~TeV probe PeV cosmic rays and astrophysical neutrino sources. Their strong attenuation on background radiation makes the observed flux sensitive to the location of the emission and to possible modifications of photon propagation. Searches for an isotropic component require rejection of an overwhelming hadronic background, but uncertainties in air-shower physics, cosmic-ray composition and detector response can lead to differences between simulated and measured events. Here we report upper limits on the isotropic diffuse gamma-ray flux from $100$~TeV to $10$~PeV using $1078$ days of Carpet-3 observations collected between 2022 and 2026. We identify photon candidates with a neural network combining the surface-array signals and measurements from the upgraded $410~\mathrm{m^2}$ muon detector. Unsupervised domain adaptation uses unlabeled experimental events during training to mitigate the impact of simulation--data discrepancies on the classification. The limits are among the strongest available and provide benchmarks for tests of extended Galactic emission and modified photon propagation from extragalactic sources.
\end{abstract}

\maketitle

\section{Introduction}
Gamma rays above $100$~TeV provide a direct test of the origin of high-energy astrophysical neutrinos, since hadronic interactions produce both particles with comparable energies and fluxes~\cite{AhlersMurase2014,KalashevTroitskyGamma}. Pair production on background radiation strongly attenuates photons from distant sources, with the absorption length falling to approximately $10$~kpc around $1$~PeV, e.g.~\cite{VernettoLipari2016}. A diffuse component extending to high Galactic latitudes could therefore reveal cosmic-ray interactions in nearby or extended Galactic structures. Cosmic rays escaping the disk may produce photons and neutrinos through interactions with circumgalactic gas, see e.g.\ \cite{KalashevMartynenkoTroitsky2023} and references therein. Alternatively, cosmic rays from a nearby accelerator may illuminate the wall of the Local Bubble, producing emission over a large fraction of the sky~\cite{BouyahiaouiKachelriessSemikoz2020}. The different distances of these emission regions imply different attenuation of their photon signals relative to the accompanying neutrinos. Searches in the sub-PeV and PeV ranges thus probe both the transport of cosmic rays beyond the Galactic disk and the local contribution to the astrophysical neutrino flux.

Diffuse emission at these energies also provides a probe of particle physics beyond the Standard Model. Decays of heavy dark-matter particles in the Galactic halo can generate photons above $100$~TeV over a broad range of arrival directions~\cite{EsmailiSerpico2021}. For extragalactic sources, Lorentz-violating modifications of pair production can reduce photon attenuation in some regions of parameter space~\cite{RubtsovSatuninSibiryakov2012}. Photon conversion into axionlike particles in magnetic fields offers another way to transport energy across optically thick regions, followed by reconversion into photons in the Milky Way. Applied to a population of neutrino sources, this mechanism can produce an observable diffuse photon component~\cite{EcknerCalore2022}. These scenarios predict emission over large angular scales, with residual anisotropy determined by the source distribution and, for axionlike particles, the Galactic magnetic field. Limits obtained under an isotropic-flux hypothesis provide an observational benchmark for such emission.

In this work, we search for an isotropic diffuse photon flux with
Carpet-3, an air-shower array at the Baksan Neutrino Observatory. We
analyze $1078$ days of observations collected between 2022 and 2026
with the upgraded muon detector of total area
$410~\mathrm{m^2}$. The search exploits the smaller muon content of
photon-induced showers compared to hadronic showers. To combine this
information with the detailed pattern of energy deposits in the central
scintillator array, we employ a neural network classifier trained on
Monte Carlo simulations.

An isotropic signal does not provide a signal-free control region
selected by arrival direction, so photon identification relies on
shower observables and their simulated distributions. Discrepancies
between simulations and measurements can therefore affect the
classification. Possible sources include uncertainties in hadronic
interaction models and in the primary cosmic-ray
composition~\cite{KuznetsovEtAl2024}, together with imperfect modeling
of the detector response. For Carpet-3, the latter could involve
unaccounted-for features of the long-serving central scintillator array
and its readout electronics. The relative contributions of these effects
to the observed discrepancies have not been established. A classifier
using the detailed detector signals may be sensitive to such mismatches
through correlations that are imperfectly reproduced in simulations.

To mitigate the impact of simulation--data discrepancies on the
classification, we use unsupervised domain
adaptation~\cite{GaninLempitsky2015}. This method combines labeled
simulated showers with unlabeled experimental events during training
to encourage features that are informative for photon--hadron
separation and less sensitive to whether an event is simulated or
measured. It addresses the effect of the discrepancies on the
classifier without requiring their physical or instrumental origin to
be specified. With this analysis, we obtain $90\%$ confidence-level
integral and differential flux limits in the energy range from
$100$~TeV to $10$~PeV. The differential limits improve on previous
Carpet-2 results in the PeV range~\cite{Dzhappuev_2023}. This paper is structured as follows: in Section \ref{sec:2} we briefly describe the Carpet-3 experiment, the measured dataset and Monte-Carlo simulated datasets; Section \ref{sec:3} describes the neural network architecture and training; Section \ref{sec:4} is dedicated to obtaining the gamma-ray flux upper limits; Section \ref{sec:concl} presents a brief summary and conclusions. 
\section{Experimental installation and datasets}
\label{sec:2}
\subsection{Carpet-3 installation and data}
Carpet-3 is an extensive air shower (EAS) surface detector array located at the Baksan Neutrino Observatory in Neutrino village, Kabardino-Balkarian Republic, Russia. The array is sensitive to cosmic rays with the primary particle energies in the range of \mbox{$\sim( 10^{14}-10^{16})$ eV}. Carpet-3 includes the central Carpet detector array, consisting of 400 liquid scintillator detectors, arranged in a 20 by 20 grid with continuous coverage of an area of $196\ \textrm{m}^2$, and an underground muon detector with plastic scintillator detectors with a total area of $410\ \textrm{m}^2$. Historically, the muon detector was built in several stages. The oldest part with an area of $175\ \textrm{m}^2$ measures the total energy release in the detectors and estimates the total number of muons detected $\textrm{n}_\mu^{175}$. The latest $235\ \textrm{m}^2$ part is operating in the counting mode, estimating the number of muons $\textrm{n}_\mu^{235}$ as a function of the number of triggered detectors. This difference is taken into account in Monte-Carlo simulations and for this reason we use both the total number of muons $\textrm{n}_\mu^{410} = \textrm{n}_\mu^{175} + \textrm{n}_\mu^{235}$ and $\textrm{n}_\mu^{175}$ as separate variables.

All events pass the following quality cuts used in the standard reconstruction pipeline:
\begin{enumerate}
    \item The reconstructed shower axis is inside the central Carpet array excluding the detectors located at the boundary of the central array,
    \item The measured energy release in the central Carpet array exceeds 5000 relativistic particles (r.p.),
    \item The number of triggered detectors of the central Carpet array exceeds 200,
    \item The reconstructed zenith angle is $\theta \leq 40^\circ$
\end{enumerate}

The dataset used in this article includes 161 076 events detected in 1078 days of operation between June 17, 2022, and April 7, 2026. For a more detailed overview of the experimental installation, data collection, reconstruction algorithms and Monte-Carlo simulations see, e.g. Refs. \cite{Szabelski_2009, Dzhappuev_2019, Dzhappuev_2007, dzhappuev2015}
\subsection{Monte-Carlo datasets}
The showers are modeled using CORSIKA 7.8000 \cite{CORSIKA} with QGSJET-III \cite{Ostapchenko:2024qgsjetIII} and FLUKA 2024.1~\cite{Ballarini:2024isa} as hadronic interaction models. The responses of the detectors are modeled using a dedicated code routine used in all previous Carpet-2 and Carpet-3 data analyses and are stored in the same format as the experimental data. The subsequent shower reconstruction algorithm is identical between the simulated and real data.

The Monte-Carlo (MC) set uses 15 000 CORSIKA showers with proton primaries and 15 000 showers with gamma-ray primaries with energies from 10~TeV to 10~PeV. Showers are modeled with the $E^{-1}$ energy spectrum to ensure that the high-energy events are well-represented in the dataset. The azimuthal-angle distribution is uniform, zenith-angle distribution is consistent with the geometric exposure. Each shower is used 100 times with a random axis offset, covering a 20 by 20 meters square centered on the central Carpet-3 detector.

From this simulation, we construct two distinct datasets, subjected to the same quality cuts as the experimental data:
\begin{enumerate}
    \item Gamma-ray and proton dataset, $E^{-1}$ primary energy spectrum, 426 981 events.
    \item An experimental-like proton dataset, in which the reconstructed shower size $N_e$ spectrum matches that of the actual experimental data, 78 911 events.
\end{enumerate}
We use the latter set, and not the complete proton dataset, for domain adaptation, as the difference in energy (and, subsequently, shower size $N_e$) spectra proved to be a critical factor during training that allowed the neural network to easily separate MC events from experimental ones. This happens because of the steeply falling spectrum - the MC data set contains many more high-energy events, and so the energy serves as a very strong distinguishing feature. 

We do not include heavier elements into the datasets as they are easier to distinguish from gamma-ray-induced showers. Moreover, the exact mass composition of cosmic rays in this energy region is still not precisely determined. Including only proton events provides a conservative dataset for the network to learn to separate gamma rays from hadronic background. 

\section{Neural network architecture and training}
\label{sec:3}
\subsection{Network Architecture}
\begin{figure*}[t]
\centering
\includegraphics[width=.82\textwidth]{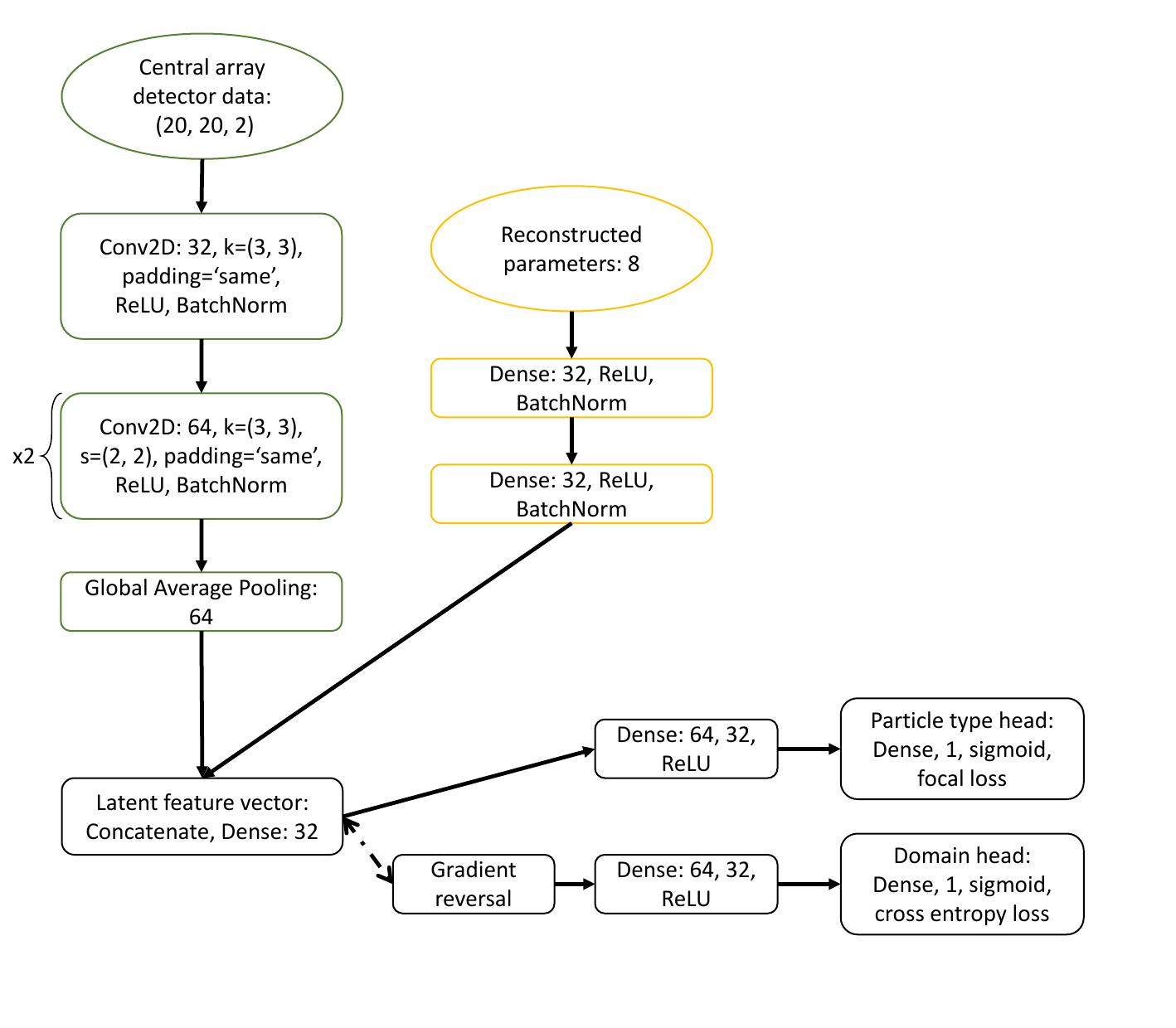}
\caption{Neural network architecture. Layer names and parameters follow Keras naming conventions.}
\label{fig:nn_arch}
\end{figure*}

We employ a dedicated neural network to predict the primary particle type of a given EAS event. It uses the energy releases in each of the 400 central Carpet scintillator detectors, which are treated as a 20 by 20 pixel image, as well as a number of reconstructed shower parameters, namely, the reconstructed shower size $N_e$, the estimated number of muons in the old muon detector $n_\mu^{175}$ and the total number of muons $n_\mu^{410}$, shower axis coordinates X, Y, reconstructed arrival direction $(\varphi, \theta)$ and a variable $C_k$ introduced in \cite{Conceicao2022}, that represents a measure of azimuthal anisotropy of the lateral particle density distribution. The network consists of  several parts (see Figure~\ref{fig:nn_arch} for complete architecture):
\begin{enumerate}
    \item A convolutional sub-network for the detector energy release data. Due to the small size of the Carpet array, each event triggers most of the detectors inside the central array, and is well-suited for 2D convolutions, avoiding the problems with sparse data. For each detector, we pass the energy release normalized to the maximum in the whole data set, and energy release normalized to the maximum in an individual event, as two separate channels. This is done to preserve both the overall energy release and the relative differences between the detectors in each event.
    \item A small two-layer fully-connected sub-network for the reconstructed features. Its output is concatenated with the convolutional sub-network's output and passed through one more fully-connected layer to form a latent feature vector of length 32. We find that increasing the number of latent features does not significantly increase the network's performance.
    \item The latent feature layer serves as input for two independent fully-connected prediction heads for the particle type and domain, with a gradient reversal layer before the domain head (see section \ref{sec:da} for details). The output of the particle type head is a number $\xi \in [0, 1]$, representing the network's confidence that the event is caused by a gamma ray.
\end{enumerate}

The network is built and trained using Keras \cite{chollet2015keras} and TensorFlow \cite{tensorflow2015-whitepaper}.

\subsection{Domain adaptation}
\label{sec:da}
When training the network using MC simulations, one should be wary that its behavior on real-world data may differ from that inferred from simulations. Although MC simulations of extensive air showers generally provide a reasonable representation of the experimental data, some variables are not perfectly modeled. This is particularly important for machine learning classifiers that use those variables as input features, as they may be more sensitive to this discrepancy than standard reconstruction algorithms. Removing those variables from the inputs altogether is not an ideal solution, because they often carry a lot of information about the events. One such example is the particle number density of the muon component, which often serves as a main way of differentiating between types of primary particles \cite{Aab_2015, Dembinski_2019}.

To mitigate this and make network predictions more reliable, we use domain adaptation with gradient reversal, first introduced in \cite{GaninLempitsky2015}. This approach involves training two sub-networks, one for particle type classification, and the other for domain discrimination between simulations and real-world data. The goal is to maximize classification performance while minimizing simulation-data discrimination. This is achieved by adding a gradient reversal layer, which multiplies the gradients passing through by $-\lambda$, before the domain discrimination head. This discourages the network from constructing latent features whose distributions differ significantly between Monte-Carlo and the real world and allows the domain prediction head to separate data domains. An ideal network would have 50 \% accuracy in domain classification, while maintaining fair particle type accuracy.
\subsection{Training results}
We train the network both on Monte-Carlo data described above and on a subsample of the experimental dataset for the purpose of domain adaptation. The particle type classifier learns only on examples from the $E^{-1}$ Monte-Carlo set, and the domain discriminator learns only on the experimental-like proton MC set and the experimental data subsample. We do not expect a significant number of gamma-ray events in the data, and so we do not use the gamma-ray MC for domain adaptation. We split the dataset into training and test sets, and evaluate the network performance on the test set after training. The test set contains 160 000 events. We also withhold a 20\% portion of the training set to use as validation data during the training process.

For the task of setting upper limits on the number of gamma-ray induced events, it is important to reduce the number of false positive identifications. We use focal loss \cite{lin2018focallosst} as a loss function for particle type classification to help achieve that, as it makes the network focus on hard-to-classify events. For domain discrimination, regular binary cross-entropy is used, with a weight factor of 0.2 that prevents it from overwhelming the classification loss.

During training, we slowly increase the influence of gradient reversal, starting from $\lambda = 0$, to make the process more stable and allow the particle-type classifier to learn important feature relationships without the domain classifier completely erasing the domain discriminating features. We also use early stopping, monitoring the classification loss on validation data with a patience of 3 epochs, due to the adversarial nature of domain adaptation. This proved to make the training results reproducible and stable.

\begin{figure*}[t]
\centering

\begin{minipage}[t]{0.48\textwidth}
    \centering
    \includegraphics[width=\linewidth]{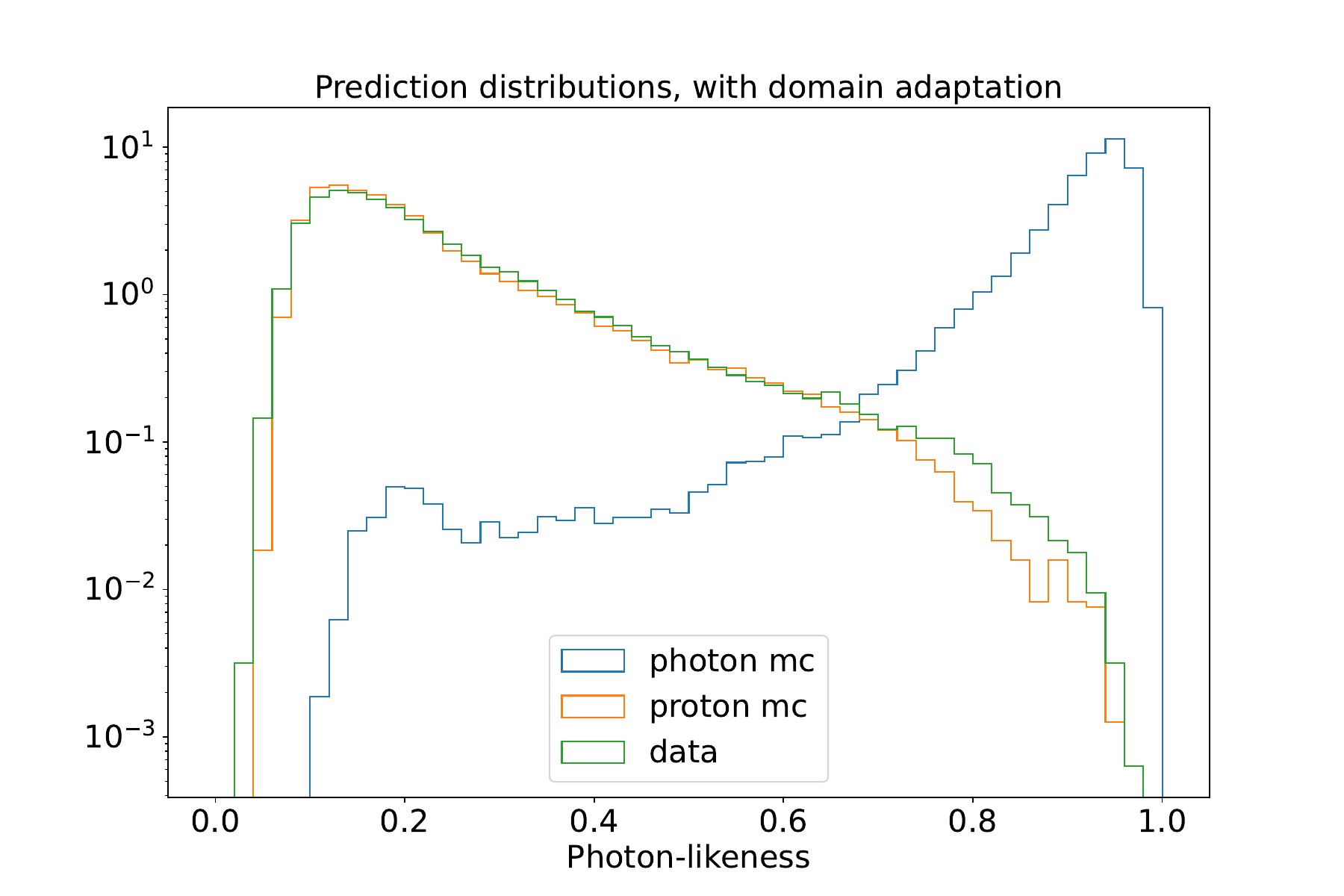}
    \caption{Prediction distributions for the MC test dataset
    and experimental data.}
    \label{fig:preds}
\end{minipage}
\hfill
\begin{minipage}[t]{0.48\textwidth}
    \centering
    \includegraphics[width=\linewidth]{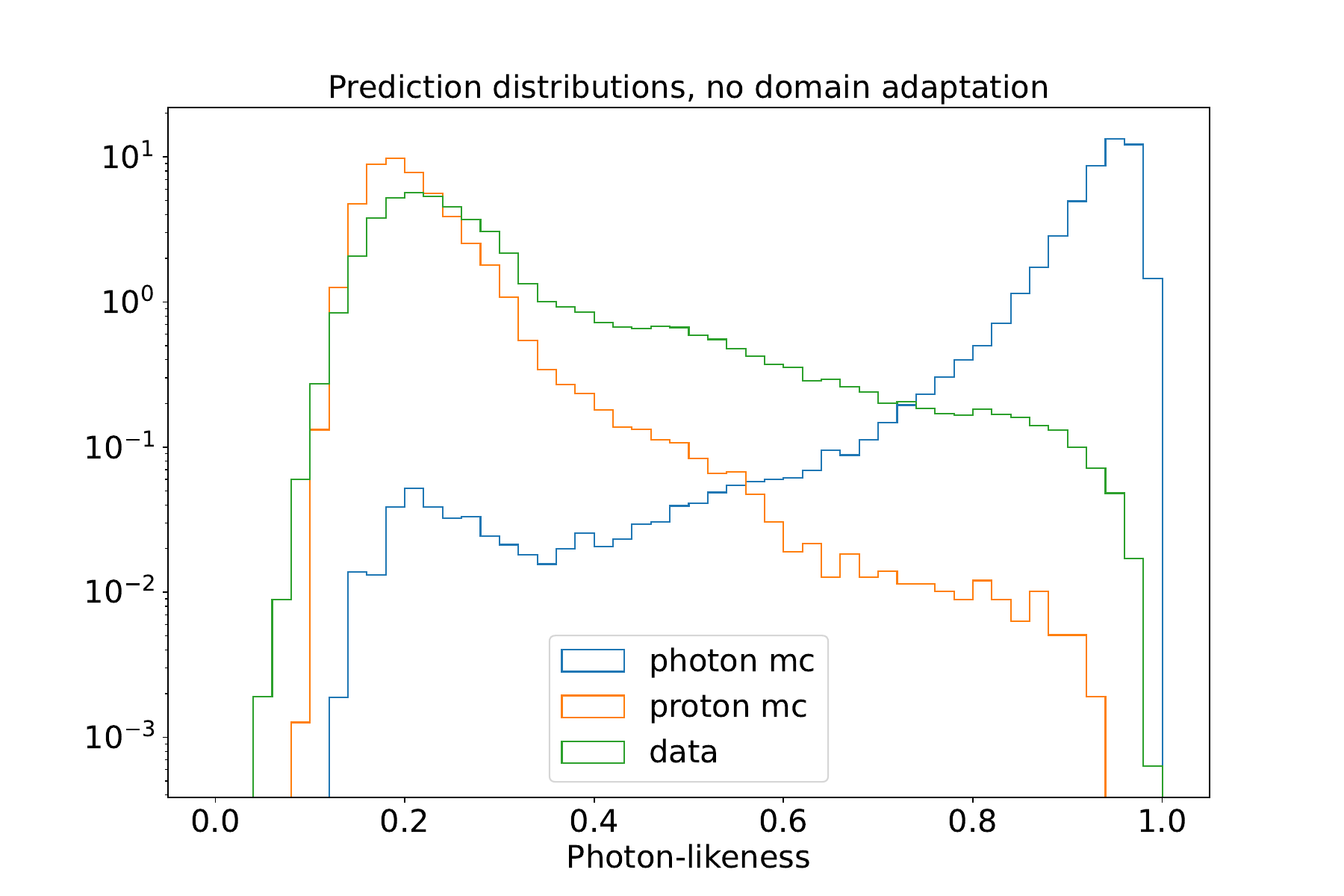}
    \caption{Prediction distributions for the MC test dataset
    and experimental data without domain adaptation.}
    \label{fig:noda_preds}
\end{minipage}
\end{figure*}

Figure \ref{fig:preds} presents the resulting prediction distribution histogram for the Monte-Carlo test set and the experimental sub-sample used in training. Note that domain adaptation brings the histograms of MC protons and experimental data to closer agreement. In fact, as experimental data contains not only protons, but also heavier primaries, we expect that the network would be less certain on those events, so the experimental data histogram should be slightly higher than MC, and that is indeed the case.

We check the effects of domain adaptation by training the same network with domain loss set to zero. The resulting prediction distributions are shown in Figure~\ref{fig:noda_preds}. 

We then calculate the KL-divergence between the predictions for experimental data and MC for both networks. The process of domain adaptation reduces it by about a factor of 15. We note the significant difference between prediction distributions for Monte-Carlo and experimental data in the case of no domain adaptation. As such, domain adaptation is crucial in making the classifier network behave reliably on real-world data. We would like to note that while such neural network classifiers are sensitive to the hadronic interaction model used for air shower simulations, as was shown in e.g. \cite{TA-gamma, Kuznetsov:2023kss}, the process of domain adaptation should greatly reduce the systematics arising from this factor, as it aligns the feature distributions between MC and experiment regardless of the hadronic interaction model. We are planning to perform additional studies of uncertainties caused by the hadronic model choice in the setting of domain adaptation in the future. 

\begin{figure*}[t]
\centering
\includegraphics[width=.48\textwidth]{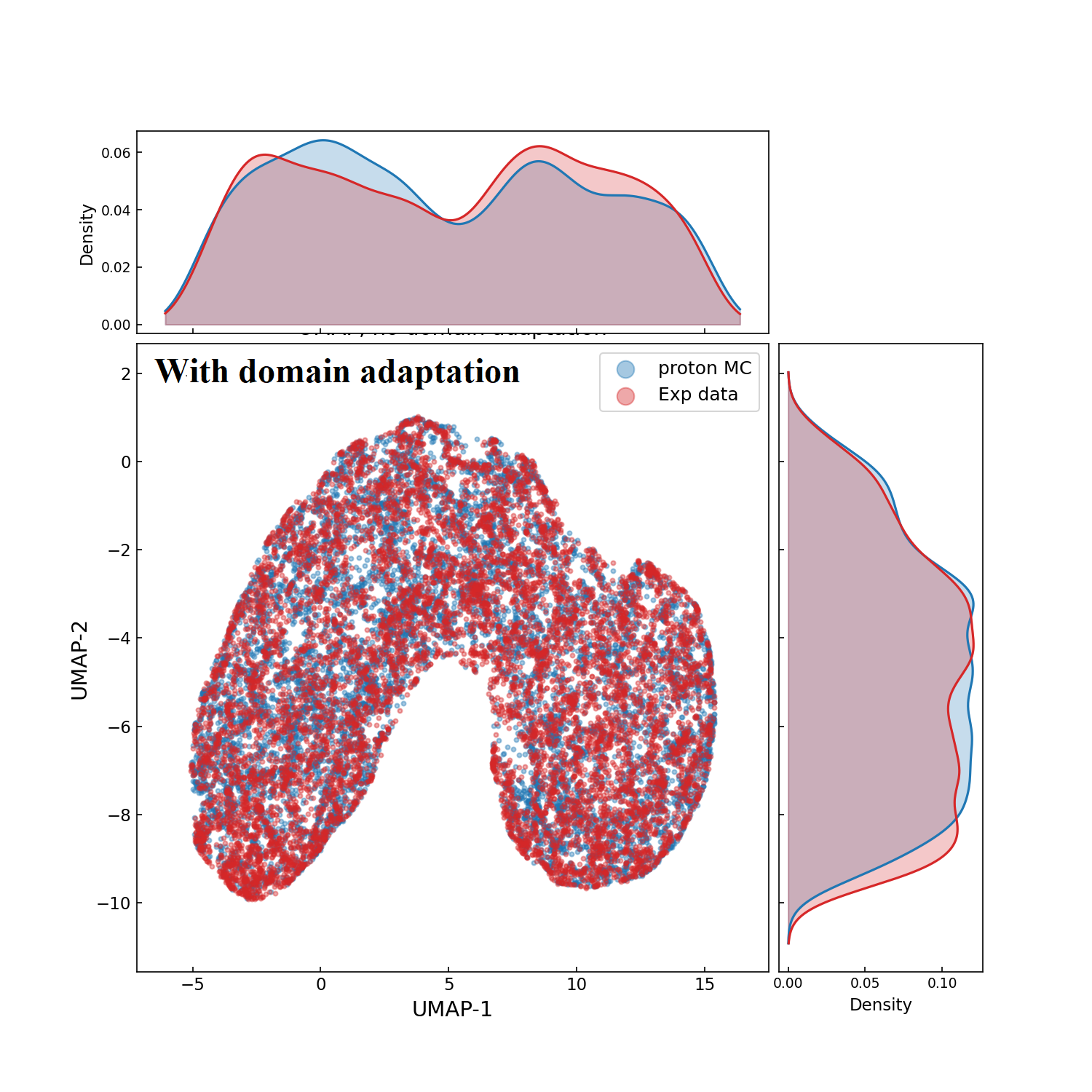}
\hfill
\includegraphics[width=.48\textwidth]{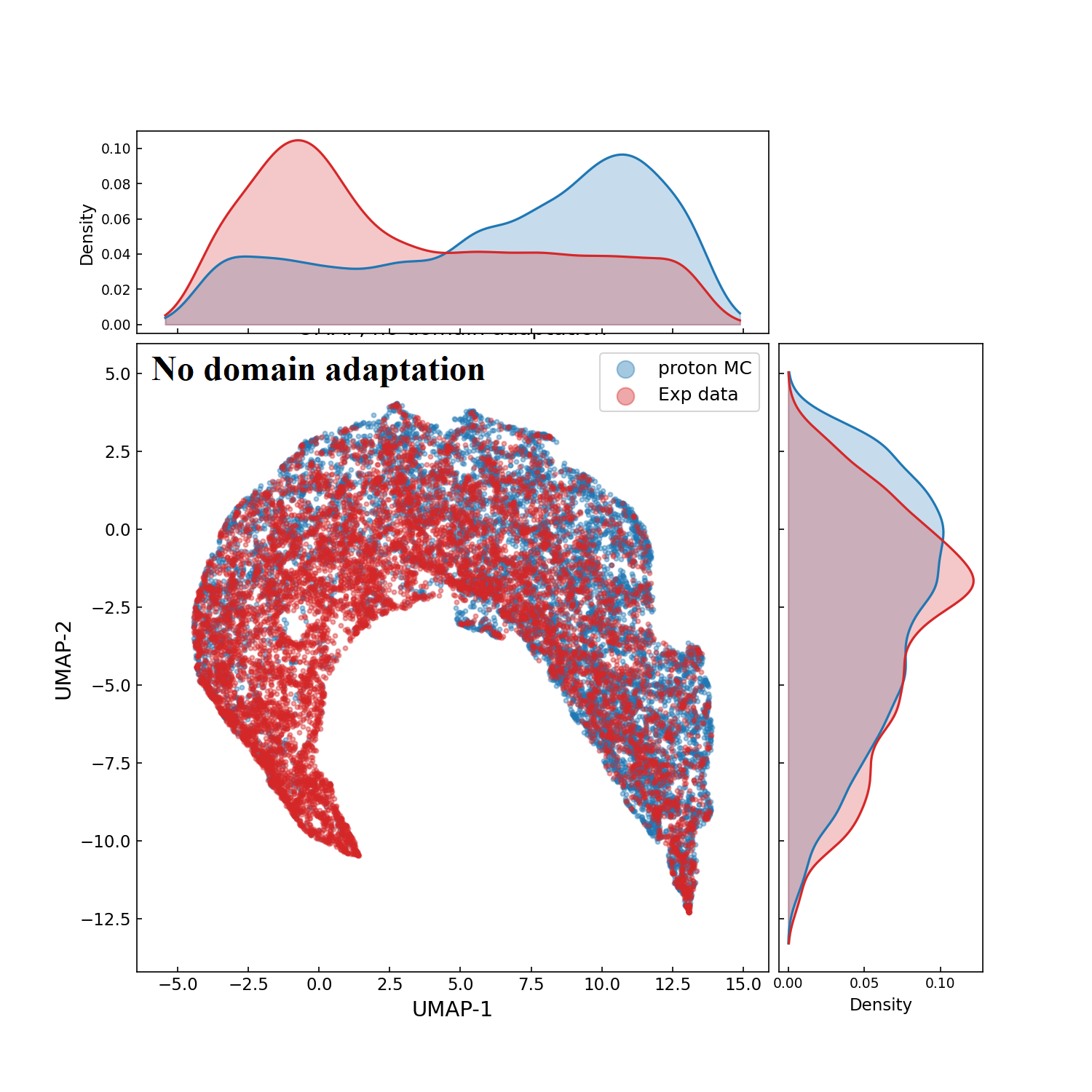}
\caption{UMAP projection of the 32-dimensional latent feature space to two dimensions for the proton set and experimental data (10,000-event subsample). Left: with domain adaptation; right: without.}
\label{fig:umap}
\end{figure*}

As a visual representation of the effects of domain adaptation, we use a UMAP projection \cite{umap}, a dimensionality reduction technique that projects the 32-dimensional latent feature space into two dimensions while aiming to preserve the local structure of the data. This allows us to visually compare the feature distributions of the two domains, with the corresponding marginal density distributions shown alongside the projection (Figure \ref{fig:umap}). Without domain adaptation, the domains form clearly distinguishable marginal distributions, indicating that the learned features retain domain-specific differences. In contrast, applying domain adaptation leads to greater overlap between the projected feature distributions, consistent with the goal of learning a more domain-invariant representation.

\section{Isotropic flux upper limits}
\label{sec:4}

To estimate the gamma-ray flux using the neural network classifier, we fix the classification threshold $\xi_0$ and identify events with $\xi > \xi_0$ as photon candidates. Since gamma rays are expected to constitute only a tiny fraction of all events, the flux upper limits are mainly determined by the network's false positive rate. To obtain the strongest differential flux limits, we maximize the quantity
\begin{equation}
\alpha = \frac{\varepsilon(\xi_0)}
{N_{\mathrm{bkg}}^{90}(\xi_0)},
\end{equation}
where $\varepsilon$ is the gamma-ray classification efficiency (true positive rate), and $N_{\mathrm{bkg}}^{90}$ is the 90\% confidence level (CL) upper limit on the expected number of background events misclassified as photons.

We estimate this number using an independent, experimental-like proton MC sample that was not used for classifier training. The number of misclassified proton events in this sample is scaled to the size of the full experimental dataset to obtain the expected background count. Its upper confidence limit, $N_{\mathrm{bkg}}^{90}$, is then determined assuming Poisson statistics and following the Feldman--Cousins approach \cite{Feldman_1998}. This assumes that our false-positive rates hold for experimental data, i.e. prediction distributions are the same for proton MC and experimental data, which is precisely the effect of domain adaptation. 

To convert the number of photon events into a differential flux, we calculate the detector effective exposure from MC simulations. For an isotropic flux, the effective exposure in an energy bin can be expressed as
\begin{equation}
\mathcal{E}(E)
= T \int_{\Omega} A_{\mathrm{eff}}(E,\theta,\phi)\,
d\Omega,
\end{equation}
where $T$ is the observation time, $\Omega$ is the solid angle of the considered field of view, and $A_{\mathrm{eff}}$ is the detector effective area, including the effects of trigger, event reconstruction, quality cuts, and photon selection with $\xi > \xi_0$. 

\begin{figure}[tbp]
\centering
\includegraphics[width=\columnwidth]{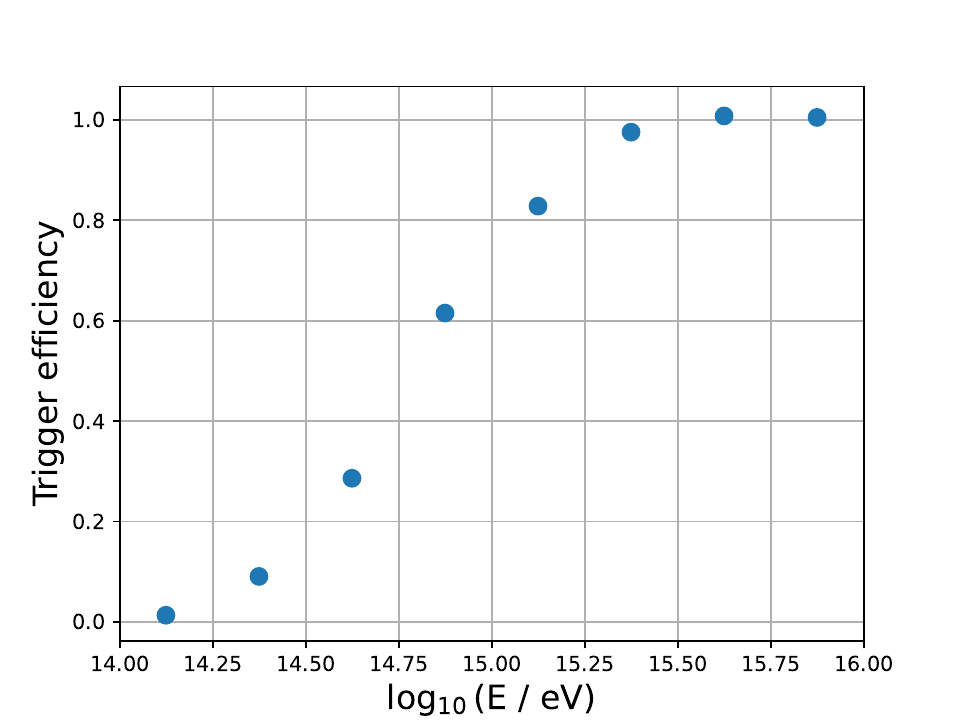}
\caption{Trigger efficiency for gamma-ray primaries. Relative statistical uncertainties are below 5\% in each bin.}
\label{fig:eff}
\end{figure}

The gamma-ray trigger and reconstruction efficiency, calculated as the ratio of the number of reconstructed MC events passing all quality cuts to the total number of generated gamma-ray MC events in each energy bin, is shown in Figure~\ref{fig:eff}.

\begin{figure*}[t]
\centering

\begin{minipage}[t]{0.48\textwidth}
    \centering
    \includegraphics[width=\linewidth]{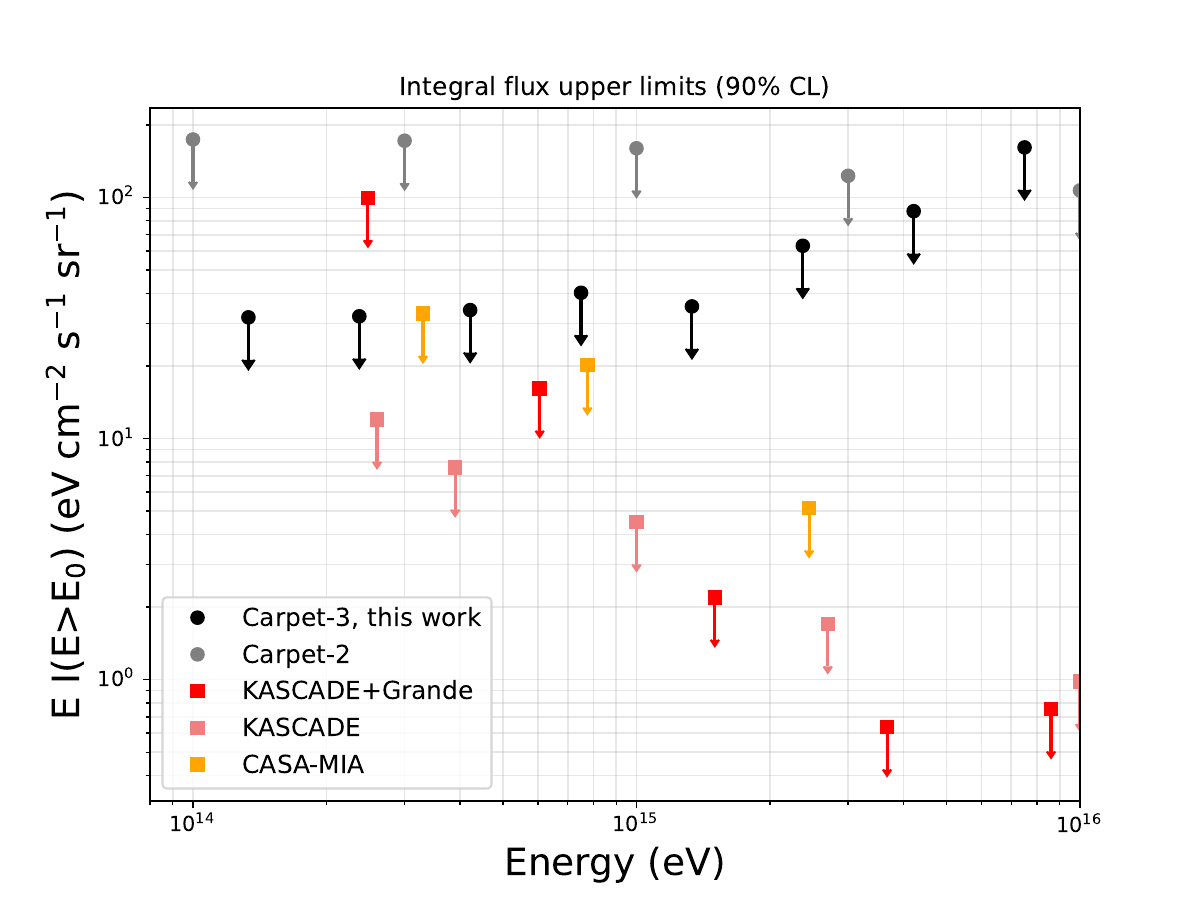}
    \caption{90\% CL integral isotropic diffuse gamma-ray flux
    upper limits. Black circles - Carpet-3, this work.
    Grey circles - Carpet-2 limits \cite{Dzhappuev_2023}.
    Red boxes - KASCADE and KASCADE-Grande \cite{Apel_2017}.
    Pink boxes - KASCADE \cite{Schatz:2003aw}.
    Orange boxes - CASA-MIA \cite{Chantell_1997}.}
    \label{fig:flux_int}
\end{minipage}
\hfill
\begin{minipage}[t]{0.48\textwidth}
    \centering
    \includegraphics[width=\linewidth]{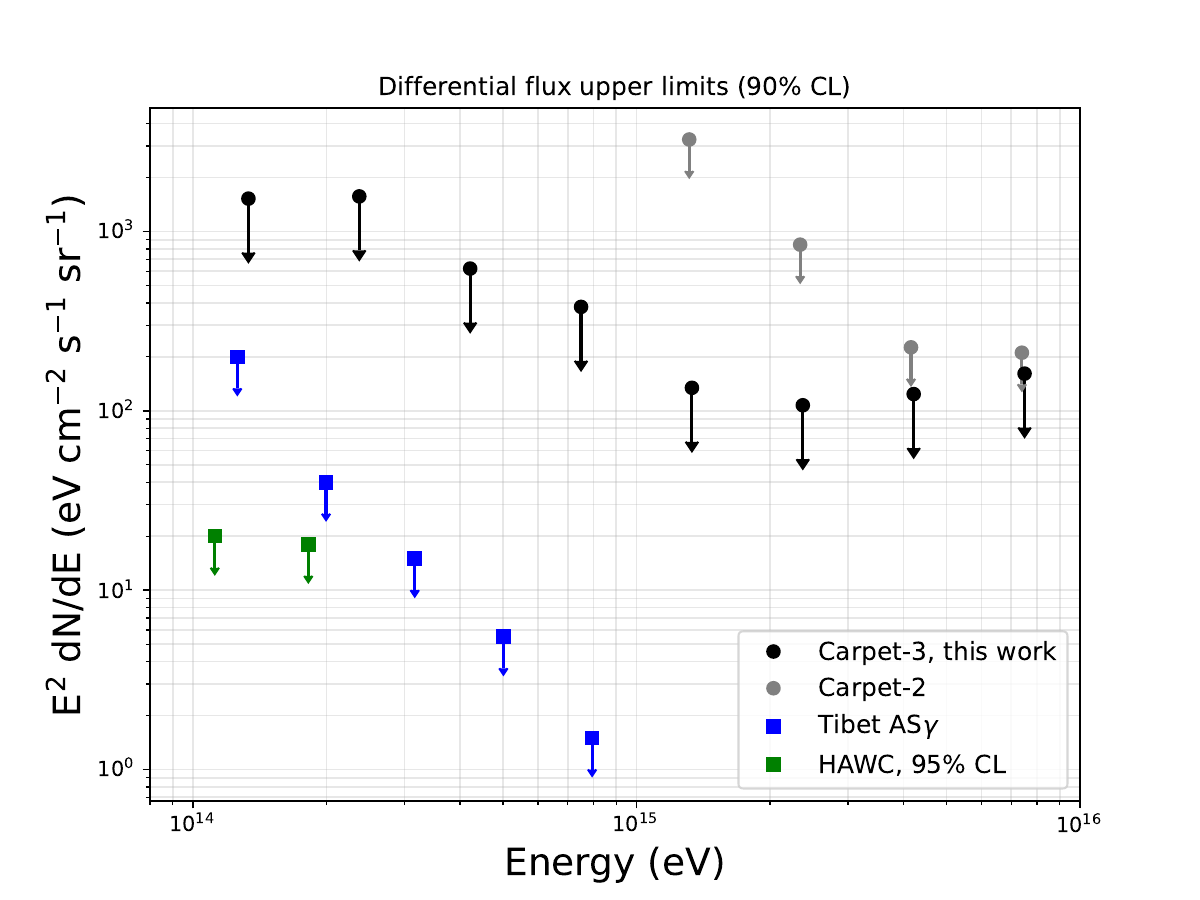}
    \caption{90\% CL differential isotropic diffuse gamma-ray
    flux upper limits. Black circles - Carpet-3, this work.
    Grey circles - Carpet-2 limits \cite{Dzhappuev_2023}.
    Green boxes - HAWC \cite{HAWC:2022uka}.
    Blue boxes - analysis of Tibet AS$\gamma$ results
    by other authors \cite{Neronov_2021ezg}.}
    \label{fig:flux_diff}
\end{minipage}

\end{figure*}

Traditionally, experiments report upper limits on the integral flux at energies above PeV. and limits or measurements of the differential flux at energies below PeV. As our working energy range is 100 TeV - 10 PeV, we report both integral and differential flux limits. We optimize the classification threshold $\xi_{0}$ in the way described above for each energy bin independently (this means ${E_{\mathrm{low}} < E < E_{\mathrm{high}}}$ for the differential bins and ${E > E_{\mathrm{low}}}$ for the integral bins). We assume $E^{-2}$ gamma-ray energy spectrum in each differential energy bin, and $E^{-2}$ spectrum across the whole energy range for the integral limits. The resulting $90\%$ CL integral and differential flux limits are given in Tables \ref{tab:flux_int} and \ref{tab:flux_diff}, and shown in Figures \ref{fig:flux_int} and \ref{fig:flux_diff} alongside the results published by other experiments. We note that the Carpet-3 upgraded muon detector and neural network classification allow us to obtain much stronger differential and integral flux limits, compared to previous Carpet-2 results, with a comparable exposure time. 

At the lower end of the energy range, the limits are significantly improved because of the stronger gamma-hadron discrimination. At the higher end, it is easier to separate the photon-induced showers from the background, and thus the limits are mainly determined by exposure time, and do not improve as much. Due to the very small trigger efficiency for the lowest energy bins, the integral limits and proton suppression (the fraction of proton events passing the classification cut) are dominated by contributions from the higher-energy bins. This happens both because of the exposure contribution and limited statistics of successfully reconstructed low-energy MC events. In the differential regime below 1 PeV we observe an excess of photon candidate events above the expected background. We do not interpret the excess as a physical gamma-ray flux due to possible data-MC discrepancies in trigger and reconstruction efficiency in the energy region close to the trigger threshold. Careful consideration of these effects is deferred to a separate study. 

We note that the isotropic diffuse differential flux upper limits above 1 PeV were previously published only by Carpet-2, and this work improves them, giving the strongest differential flux limits in this energy range. The integral limit at $\mathrm{E > 100\ TeV}$ is also the most stringent published one. A limit on the fraction of gamma-rays in this energy range was published by the GRAPES-3 experiment \cite{GRAPES-3-gamma}, but conversion from gamma-ray fraction to gamma-ray flux is not straightforward.

\section{Conclusions}
\label{sec:concl}
In this paper, we present new upper limits on the isotropic diffuse gamma-ray flux obtained from Carpet-3 experimental data collected between June 2022 and April 2026. By leveraging the recently upgraded muon detector and employing a neural network-based classifier with unsupervised domain adaptation, we have significantly enhanced the experiment's sensitivity to gamma-ray-induced extensive air showers compared to the previous Carpet-2 analysis. Domain adaptation was crucial to make the network predictions reliable when using it on experimental data and mitigate the systematic uncertainties arising from Monte-Carlo imperfections.

We observe an excess of photon candidate events below 1 PeV above the expected background. However, converting this excess into a gamma-ray flux measurement requires further dedicated analysis, which we plan to address in a future work.
\\

\section{Acknowledgments}

We gratefully acknowledge I. Kharuk for valuable insight and discussions.

This work was supported by the Russian Science Foundation project No. 25-12-00333.

\begin{table*}[t]
\centering
\caption{90\% CL integral gamma-ray flux limits. $E^{-2}$ spectrum is assumed. $\varepsilon_\gamma$ - gamma-ray detection efficiency, taking into account both trigger and neural network efficiency. $\mathrm{\Phi_\gamma}$ - integral flux upper limits.
\label{tab:flux_int}}
\smallskip
\resizebox{\textwidth}{!}{%
\begin{tabular}{|c|c|c|c|c|c|}
\hline
\makecell{$E_{\rm min}$, \\ $\log_{10}(E / \textrm{eV})$} 
& $\varepsilon_\gamma$
& \makecell{Proton \\ suppression} 
& \makecell{Expected \\ background} 
& \makecell{Observed photon \\ candidates} 
& \makecell{$\mathrm{\Phi_\gamma},\ \textrm{cm}^{-2} \cdot \textrm{s}^{-1} \cdot \textrm{sr}^{-1}$} \\
\hline
14    & 0.004 & $< 1.3 \cdot 10^{-5}$ & 2   & 1 & $ < 3.2 \cdot 10^{-13}$\\
14.25 & 0.03  & $< 1.3 \cdot 10^{-5}$ & 2   & 1 & $ < 1.8 \cdot 10^{-13}$\\
14.5  & 0.09  & $< 1.3 \cdot 10^{-5}$ & 1.9 & 1 & $ < 1.1 \cdot 10^{-13}$\\
14.75 & 0.21  & $< 1.4 \cdot 10^{-5}$ & 1.6 & 1 & $ < 7.2 \cdot 10^{-14}$\\
15    & 0.28  & $< 1.8 \cdot 10^{-5}$ & 1.3 & 0 & $ < 3.54 \cdot 10^{-14}$\\
15.25 & 0.36  & $< 4 \cdot 10^{-5}$   & 0.9 & 0 & $ < 3.53 \cdot 10^{-14}$\\
15.5  & 0.54  & $< 2 \cdot 10^{-4}$   & 0.9 & 0 & $ < 2.8 \cdot 10^{-14}$\\
15.75 & 0.84  & $< 1 \cdot 10^{-3}$   & 0.8 & 0 & $ < 2.8 \cdot 10^{-14}$\\
\hline
\end{tabular}
}
\end{table*}

\begin{table*}[t]
\centering
\caption{90\% CL differential gamma-ray flux limits. $E^{-2}$ spectrum is assumed. $\varepsilon_\gamma$ - gamma-ray detection efficiency, taking into account both trigger and neural network efficiency. $\mathrm{dN/dE}$ - differential flux upper limits.\label{tab:flux_diff}}
\smallskip
\resizebox{\textwidth}{!}{%
\begin{tabular}{|c|c|c|c|c|c|}
\hline
\makecell{Bin center, \\ $\log_{10}(E / \textrm{eV})$} 
& $\varepsilon_\gamma$ 
& \makecell{Proton \\ suppression} 
& \makecell{Expected \\ background} 
& \makecell{Observed photon \\ candidates} 
& \makecell{$\mathrm{E^2 \cdot dN/dE}$, \\ $\textrm{eV} \cdot \textrm{cm}^{-2} \cdot \textrm{s}^{-1} \cdot \textrm{sr}^{-1}$} \\
\hline
14.125 & 0.01 & $6\cdot 10^{-3}$ & 15  & 11 & $ < 1.53 \cdot 10^{3}$\\
14.375 & 0.08 & $6\cdot 10^{-4}$ & 8  & 55 & $ < 1.57 \cdot 10^{3}$\\
14.625 & 0.19 & $2\cdot 10^{-4}$ & 5  & 27 & $ < 6.2 \cdot 10^{2}$\\
14.875 & 0.38 & $7\cdot 10^{-5}$ & 3 & 18 & $ < 3.8 \cdot 10^{2}$\\
15.125 & 0.42 & $4\cdot 10^{-5}$ & 1.6 & 2 & $ < 1.34 \cdot 10^{2}$\\
15.375 & 0.40 & $4\cdot 10^{-5}$ & 0.9 & 0 & $ < 1.07 \cdot 10^{2}$\\
15.625 & 0.62 & $2\cdot 10^{-4}$ & 0.9 & 0 & $ < 1.24 \cdot 10^{2}$\\
15.875 & 0.84 & $1\cdot 10^{-3}$ & 0.8 & 0 & $ < 1.61 \cdot 10^{2} $\\
\hline
\end{tabular}%
}
\end{table*}

\clearpage
\bibliographystyle{apsrev4-2}
\bibliography{biblio}

@article{Conceicao2022,
doi = {10.1088/1475-7516/2022/10/086},
url = {https://doi.org/10.1088/1475-7516/2022/10/086},
year = {2022},
month = {oct},
publisher = {IOP Publishing},
volume = {2022},
number = {10},
pages = {086},
author = {Conceição, R. and Gibilisco, L. and Pimenta, M. and Tomé, B.},
title = {Gamma/hadron discrimination at high energies through the azimuthal fluctuations of air shower particle distributions at the ground},
journal = {JCAP},
}

@techreport{CORSIKA,
  author      = {Heck, D. and Knapp, J. and Capdevielle, J. N. and Schatz, G. and Thouw, T.},
  title       = {{CORSIKA: A Monte Carlo code to simulate extensive air showers}},
  institution = {Forschungszentrum Karlsruhe},
  number      = {FZKA-6019},
  year        = {1998},
  doi         = {10.5445/IR/270043064},
  url         = {https://publikationen.bibliothek.kit.edu/270043064}
}

@misc{chollet2015keras,
  title={Keras},
  author={Chollet, Fran\c{c}ois and others},
  year={2015},
  howpublished={\url{https://keras.io}},
}

@misc{lin2018focallosst,
      title={Focal Loss for Dense Object Detection}, 
      author={Tsung-Yi Lin and Priya Goyal and Ross Girshick and Kaiming He and Piotr Dollár},
      year={2018},
      eprint={1708.02002},
      archivePrefix={arXiv},
      primaryClass={cs.CV},
      url={https://arxiv.org/abs/1708.02002}, 
}

@misc{tensorflow2015-whitepaper,
title={ {TensorFlow}: Large-Scale Machine Learning on Heterogeneous Systems},
url={https://www.tensorflow.org/},
note={Software available from tensorflow.org},
author={
    Mart\'{i}n~Abadi and others},
  year={2015},
}

@misc{umap,
      title={UMAP: Uniform Manifold Approximation and Projection for Dimension Reduction}, 
      author={Leland McInnes and John Healy and James Melville},
      year={2020},
      eprint={1802.03426},
      archivePrefix={arXiv},
      primaryClass={stat.ML},
      url={https://arxiv.org/abs/1802.03426}, 
}

@article{Dzhappuev_2023,
   title={Upper Limits on the Isotropic Diffuse Flux of Cosmic PeV Photons from Carpet-2 Observations},
   volume={117},
   ISSN={1090-6487},
   url={http://dx.doi.org/10.1134/S0021364022603244},
   DOI={10.1134/s0021364022603244},
   number={3},
   journal={JETP Letters},
   publisher={Pleiades Publishing Ltd},
   author={Dzhappuev, D. D. and others},
   year={2023},
   month=Jan, pages={184–188} }

@article{Feldman_1998,
   title={Unified approach to the classical statistical analysis of small signals},
   volume={57},
   ISSN={1089-4918},
   url={http://dx.doi.org/10.1103/PhysRevD.57.3873},
   DOI={10.1103/physrevd.57.3873},
   number={7},
   journal={Physical Review D},
   publisher={American Physical Society (APS)},
   author={Feldman, Gary J. and Cousins, Robert D.},
   year={1998},
   month=Apr, pages={3873–3889} }

@article{TA-gamma,
    author = "Abbasi, R. U. and others",
    collaboration = "Telescope Array",
    title = "{Searching for EeV photons with Telescope Array surface detector and neural networks}",
    eprint = "2512.01638",
    archivePrefix = "arXiv",
    primaryClass = "astro-ph.GA",
    doi = "10.1088/1475-7516/2026/06/052",
    journal = "JCAP",
    volume = "06",
    pages = "052",
    year = "2026"
}

@article{Neronov_2021ezg,
    author = "Neronov, A. and Semikoz, D. and Vovk, Ie.",
    title = "{New limit on high Galactic latitude PeV {\ensuremath{\gamma}}-ray flux from Tibet AS{\ensuremath{\gamma}} data}",
    eprint = "2107.06541",
    archivePrefix = "arXiv",
    primaryClass = "astro-ph.HE",
    doi = "10.1051/0004-6361/202141800",
    journal = "Astron. Astrophys.",
    volume = "653",
    pages = "L4",
    year = "2021"
}

@misc{HAWC:2022uka,
    author = {Albert, A. and others},
    collaboration = {HAWC},
    title = {{Limits on the Diffuse Gamma-Ray Background above 10 TeV with HAWC}},
    eprint = {2209.08106},
    archivePrefix = {arXiv},
    primaryClass = {astro-ph.HE},
    reportNumber = {LA-UR-22-24242},
    year = {2022},
    month = sep
}

@article{Kuznetsov:2023kss,
    author = "Kuznetsov, M. Yu. and Petrov, N. A. and Plokhikh, I. A. and Sotnikov, V. V.",
    title = "{Methods of machine learning for the analysis of cosmic rays mass composition with the KASCADE experiment~data}",
    eprint = "2311.06893",
    archivePrefix = "arXiv",
    primaryClass = "astro-ph.HE",
    doi = "10.1088/1748-0221/19/01/P01025",
    journal = "JINST",
    volume = "19",
    number = "01",
    pages = "P01025",
    year = "2024"
}

@article{Apel_2017,
   title={KASCADE-Grande Limits on the Isotropic Diffuse Gamma-Ray Flux between 100 TeV and 1 EeV},
   volume={848},
   ISSN={1538-4357},
   url={http://dx.doi.org/10.3847/1538-4357/aa8bb7},
   DOI={10.3847/1538-4357/aa8bb7},
   number={1},
   journal={The Astrophysical Journal},
   publisher={American Astronomical Society},
   author={Apel, W. D. and others},
   year={2017},
   month=Oct, pages={1} }

@article{Chantell_1997,
   title={Limits on the Isotropic Diffuse Flux of Ultrahigh Energy gamma-radiation},
   volume={79},
   ISSN={1079-7114},
   url={http://dx.doi.org/10.1103/PhysRevLett.79.1805},
   DOI={10.1103/physrevlett.79.1805},
   number={10},
   journal={Physical Review Letters},
   publisher={American Physical Society (APS)},
   author={Chantell, M. C. and others},
   year={1997},
   month={Sept}, 
   pages={1805–1808} 
}

@inproceedings{Schatz:2003aw,
    author = "Schatz, G. and others",
    title = "{Search for extremely high energy gamma rays with the KASCADE experiment}",
    booktitle = "{28th International Cosmic Ray Conference}",
    reportNumber = "FZKA-6890I",
    pages = "2293--2296",
    month = "9",
    year = "2003"
}

@article{Szabelski_2009,
   title={Carpet-3 - a new experiment to study the primary composition around the knee},
   volume={196},
   ISSN={0920-5632},
   url={http://dx.doi.org/10.1016/j.nuclphysbps.2009.09.069},
   DOI={10.1016/j.nuclphysbps.2009.09.069},
   journal={Nuclear Physics B - Proceedings Supplements},
   publisher={Elsevier BV},
   author={Szabelski, J.},
   year={2009},
   month=Dec, pages={371–374} }

@article{Dzhappuev_2019,
   title={Carpet-2 Search for PeV Gamma Rays Associated with IceCube High-Energy Neutrino Events},
   volume={109},
   ISSN={1090-6487},
   url={http://dx.doi.org/10.1134/S0021364019040015},
   DOI={10.1134/s0021364019040015},
   number={4},
   journal={JETP Letters},
   publisher={Pleiades Publishing Ltd},
   author={Dzhappuev, D. D. and others},
   year={2019},
   month=Feb, pages={226–231} }

@article{Dzhappuev_2007,
       author = {{Dzhappuev}, D.~D. and others},
        title = "{Modernization of the Carpet-2 array of the Baksan Neutrino Observatory}",
      journal = {Bulletin of the Russian Academy of Sciences, Physics},
         year = 2007,
        month = apr,
       volume = {71},
       number = {4},
        pages = {525-527},
          doi = {10.3103/S1062873807040260},
       adsurl = {https://ui.adsabs.harvard.edu/abs/2007BRASP..71..525D}
}

@misc{dzhappuev2015,
      title={Search for cosmic gamma rays with the Carpet-2 extensive air shower array}, 
      author={D. D. Dzhappuev and V. B. Petkov and A. U. Kudzhaev and N. F. Klimenko and A. S. Lidvansky and S. V. Troitsky},
      year={2015},
      eprint={1511.09397},
      archivePrefix={arXiv},
      primaryClass={astro-ph.HE},
      url={https://arxiv.org/abs/1511.09397}, 
}

@article{Ostapchenko:2024qgsjetIII,
    author = {Ostapchenko, Sergey},
    title = {{QGSJET-III model of high energy hadronic interactions: The formalism}},
    journal = {Phys. Rev. D},
    volume = {109},
    issue = {3},
    pages = {034002},
    numpages = {23},
    year = {2024},
    month = {Feb},
    publisher = {American Physical Society},
    doi = {10.1103/PhysRevD.109.034002},
    url = {https://link.aps.org/doi/10.1103/PhysRevD.109.034002},
    eprint = {2401.06202},
    archivePrefix = {arXiv},
    primaryClass = {hep-ph}
}

@article{Ballarini:2024isa,
    author = "Ballarini, Francesca and others",
    title = "{The FLUKA code: Overview and new developments}",
    doi = "10.1051/epjn/2024015",
    journal = "EPJ Nuclear Sci. Technol.",
    volume = "10",
    pages = "16",
    year = "2024"
}

@article{Dembinski_2019,
   title={Report on Tests and Measurements of Hadronic Interaction Properties with Air Showers},
   volume={210},
   ISSN={2100-014X},
   url={http://dx.doi.org/10.1051/epjconf/201921002004},
   DOI={10.1051/epjconf/201921002004},
   journal={EPJ Web of Conferences},
   publisher={EDP Sciences},
   author={Dembinski, H.P. and others},
   editor={Lhenry-Yvon, I. and Biteau, J. and Deligny, O. and Ghia, P.},
   year={2019},
   pages={02004} }

@article{Aab_2015,
   title={Muons in air showers at the Pierre Auger Observatory: Mean number in highly inclined events},
   volume={91},
   ISSN={1550-2368},
   url={http://dx.doi.org/10.1103/PhysRevD.91.032003},
   DOI={10.1103/physrevd.91.032003},
   number={3},
   journal={Physical Review D},
   publisher={American Physical Society (APS)},
   author={Aab, A. and others},
   year={2015},
   month=Feb }

@article{GRAPES-3-gamma,
    author = "Pant, B. P. and others",
    title = "{Diffuse Gamma-Rays at~10-300 TeV with~the~GRAPES-3 Experiment}",
    doi = "10.1007/978-981-97-0289-3_84",
    journal = "Springer Proc. Phys.",
    volume = "304",
    pages = "384--388",
    year = "2024"
}

@article{AhlersMurase2014,
  author        = {Ahlers, Markus and Murase, Kohta},
  title         = {Probing the {Galactic} origin of the {IceCube} excess with gamma rays},
  journal       = {Phys. Rev. D},
  volume        = {90},
  pages         = {023010},
  year          = {2014},
  doi           = {10.1103/PhysRevD.90.023010},
  eprint        = {1309.4077},
  archivePrefix = {arXiv},
  primaryClass  = {astro-ph.HE},
  url           = {https://doi.org/10.1103/PhysRevD.90.023010}
}

@article{VernettoLipari2016,
  author        = {Vernetto, Silvia and Lipari, Paolo},
  title         = {Absorption of very high energy gamma rays in the {Milky Way}},
  journal       = {Phys. Rev. D},
  volume        = {94},
  pages         = {063009},
  year          = {2016},
  doi           = {10.1103/PhysRevD.94.063009},
  eprint        = {1608.01587},
  archivePrefix = {arXiv},
  primaryClass  = {astro-ph.HE},
  url           = {https://doi.org/10.1103/PhysRevD.94.063009}
}

@article{KalashevMartynenkoTroitsky2023,
  author        = {Kalashev, Oleg and Martynenko, Nickolay and Troitsky, Sergey},
  title         = {On the contribution of cosmic-ray interactions in the
                   circumgalactic gas to the observed high-energy neutrino flux},
  journal       = {JCAP},
  volume        = {03},
  pages         = {053},
  year          = {2023},
  doi           = {10.1088/1475-7516/2023/03/053},
  eprint        = {2207.12458},
  archivePrefix = {arXiv},
  primaryClass  = {astro-ph.HE},
  url           = {https://doi.org/10.1088/1475-7516/2023/03/053}
}

@article{BouyahiaouiKachelriessSemikoz2020,
  author        = {Bouyahiaoui, M. and Kachelrie{\ss}, M. and Semikoz, D. V.},
  title         = {High-energy neutrinos from cosmic ray interactions in the
                   {Local Bubble}},
  journal       = {Phys. Rev. D},
  volume        = {101},
  pages         = {123023},
  year          = {2020},
  doi           = {10.1103/PhysRevD.101.123023},
  eprint        = {2001.00768},
  archivePrefix = {arXiv},
  primaryClass  = {astro-ph.HE},
  url           = {https://doi.org/10.1103/PhysRevD.101.123023}
}

@article{EsmailiSerpico2021,
  author        = {Esmaili, Arman and Serpico, Pasquale D.},
  title         = {First implications of {Tibet AS}$\gamma$ data for heavy dark matter},
  journal       = {Phys. Rev. D},
  volume        = {104},
  pages         = {L021301},
  year          = {2021},
  doi           = {10.1103/PhysRevD.104.L021301},
  eprint        = {2105.01826},
  archivePrefix = {arXiv},
  primaryClass  = {hep-ph},
  url           = {https://doi.org/10.1103/PhysRevD.104.L021301}
}

@article{RubtsovSatuninSibiryakov2012,
  author        = {Rubtsov, Grigory and Satunin, Petr and Sibiryakov, Sergey},
  title         = {Calculation of cross sections in {Lorentz}-violating theories},
  journal       = {Phys. Rev. D},
  volume        = {86},
  pages         = {085012},
  year          = {2012},
  doi           = {10.1103/PhysRevD.86.085012},
  eprint        = {1204.5782},
  archivePrefix = {arXiv},
  primaryClass  = {hep-ph},
  url           = {https://doi.org/10.1103/PhysRevD.86.085012}
}

@article{EcknerCalore2022,
  author        = {Eckner, Christopher and Calore, Francesca},
  title         = {First constraints on axionlike particles from {Galactic}
                   sub-{PeV} gamma rays},
  journal       = {Phys. Rev. D},
  volume        = {106},
  pages         = {083020},
  year          = {2022},
  doi           = {10.1103/PhysRevD.106.083020},
  eprint        = {2204.12487},
  archivePrefix = {arXiv},
  primaryClass  = {astro-ph.HE},
  url           = {https://doi.org/10.1103/PhysRevD.106.083020}
}

@article{KalashevTroitskyGamma,
    author = "Kalashev, Oleg E. and Troitsky, Sergey V.",
    title = "{IceCube astrophysical neutrinos without a spectral cutoff and $10^{15}${\textendash}$10^{17}$ eV cosmic gamma radiation}",
    eprint = "1410.2600",
    archivePrefix = "arXiv",
    primaryClass = "astro-ph.HE",
    reportNumber = "INR-TH-2014-023",
    doi = "10.1134/S0021364014240072",
    journal = "Pisma Zh. Eksp. Teor. Fiz.",
    volume = "100",
    number = "12",
    pages = "865--869",
    year = "2014"
}

@article{KuznetsovEtAl2024,
  author        = {Kuznetsov, M. Yu. and Petrov, N. A. and
                   Plokhikh, I. A. and Sotnikov, V. V.},
  title         = {Energy spectra of elemental groups of cosmic rays with
                   the {KASCADE} experiment data and machine learning},
  journal       = {JCAP},
  volume        = {05},
  pages         = {125},
  year          = {2024},
  doi           = {10.1088/1475-7516/2024/05/125},
  eprint        = {2312.08279},
  archivePrefix = {arXiv},
  primaryClass  = {astro-ph.HE},
  url           = {https://doi.org/10.1088/1475-7516/2024/05/125}
}

@inproceedings{GaninLempitsky2015,
  author        = {Ganin, Yaroslav and Lempitsky, Victor},
  title         = {Unsupervised Domain Adaptation by Backpropagation},
  booktitle     = {Proceedings of the 32nd International Conference on
                   Machine Learning},
  series        = {Proceedings of Machine Learning Research},
  volume        = {37},
  pages         = {1180--1189},
  year          = {2015},
  publisher     = {PMLR},
  eprint        = {1409.7495},
  archivePrefix = {arXiv},
  primaryClass  = {stat.ML},
  url           = {https://proceedings.mlr.press/v37/ganin15.html}
}
\end{document}